\documentclass[12pt]{article}

\usepackage{epsfig}

\usepackage{amssymb}

\def\be{\begin{equation}}
\def\ee{\end{equation}}
\def\ba{\begin{array}{c}}
\def\ea{\end{array}}

\newcommand{\bea}{\begin{eqnarray}}
\newcommand{\eea}{\end{eqnarray}}

\begin{document}

\begin{center}

{\Large \bf

Hermitian--non-Hermitian interfaces in quantum theory

  }

\vspace{9mm}

{Miloslav Znojil}

\vspace{9mm}

Nuclear Physics Institute CAS, 250 68 \v{R}e\v{z}, Czech Republic

{znojil@ujf.cas.cz}

{http://gemma.ujf.cas.cz/$\sim$znojil/}

\end{center}

\vspace{9mm}


\section*{Abstract}

In the global framework of quantum theory the individual quantum
systems seem clearly separated into two families with the respective
manifestly Hermitian and hiddenly Hermitian operators of their
Hamiltonian. In the light of certain preliminary studies these two
families seem to have an empty overlap. In this paper we demonstrate
that it is not so. We are going to show that whenever the
interaction potentials are chosen weakly nonlocal, the separation of
the two families may disappear. The overlaps {\it alias} interfaces
between the Hermitian and non-Hermitian descriptions of a unitarily
evolving quantum system in question may become non-empty. This
assertion will be illustrated via a few analytically solvable
elementary models.

\newpage

\section{Introduction}

In virtually any representation of quantum theory the states can be
perceived as constructed in a suitable user-friendly Hilbert space
${\cal H}$. By a number of authors \cite{Dyson,Geyer,ali,Carl} it
has been recommended to enhance the flexibility of the formalism by
making use of an {\it ad hoc}, quantum-system-adapted physical inner
product in ${\cal H}$, i.e., by an introduction of a nontrivial,
stationary metric operator $\Theta\neq \Theta(t)$. All of the other,
relevant ``physical'' operators $\Lambda$ of the observables in
${\cal H}$ (i.e., say, $\Lambda_1=Q$ representing a coordinate, or
$\Lambda_2=H$ representing the energy, etc) must be then chosen, in
Diedonn\'e's terminology \cite{Dieudonne}, quasi-Hermitian,
 \be
 \Lambda^\dagger \Theta=\Theta \Lambda\,.
 \ee
These observables become Hermitian if and only if we reach the
conventional textbook limit with $\Theta \to I$. Otherwise, our
candidates for the observables  remain manifestly non-Hermitian in
our friendly Hilbert space ${\cal H}$. The latter space (with
artificial $\Theta=I$) must be declared, therefore, auxiliary and
unphysical, ${\cal H} \to {\cal H}^{(unphysical)}$. Only the
re-incorporation of the amended metric will reinstall the space as
physical,  ${\cal H} \to {\cal H}^{(redefined)}$.

In certain very promising recent high-energy physics applications of
the formalism, say, in neutrino physics \cite{AB,ABb} people usually
restrict attention to the special form of $\Theta={\cal PC}$ where
${\cal P}$ is parity while ${\cal C}$ denotes charge. In such a
setting the stationarity of the theory represents a serious obstacle
for experimentalists, mainly because the adiabatic changes and
tuning of the parameter-dependence of the observables may lead to
multiple counterintuitive no-go theorems \cite{ali,Milburn,PLB}. At
the same time, the new degree of the kinematical freedom represented
by the nontrivial metric $\Theta \neq I$ may find its efficient use,
say, in the manipulations leading to the experimental realizations
of various quantum phase transitions in the theory
\cite{AB,BB,Denis} as well as in the laboratory \cite{Makris}.

The consistent mathematical formulation of the theories with
innovative $\Theta \neq I$ and traditional $H = T+V(x)$ proved truly
challenging \cite{book}. In practice, the main source of
difficulties can be seen in the ``smearing'' feature of the use of
generic $\Theta \neq I$ \cite{smeared}. Hugh Jones noticed that ``we
have to start with $x$'' (i.e., with $\Theta \neq I$) ``because that
is how the potential is defined'' \cite{preinterface}. His analysis
was then aimed at the search for natural interfaces ({\it alias\,}
operational connections) between the hypothetical non-Hermitian
dynamics (using $\Theta\neq I$) and the available experimental
setups (at $\Theta=I$).

In a way based on a detailed study of certain overrestricted family
of models (for purely technical reasons the interaction potentials
were kept local), Jones arrived, not too surprisingly, at a heavily
sceptical conclusion that the theory cannot be unitary. In his own
words ``the only satisfactory resolution of the dilemma is to treat
the non-Hermitian potential as an effective one, and [to] work in
the standard framework of quantum mechanics, accepting that this
effective potential may well involve the loss of unitarity''
\cite{interface}.

The Jones' conclusions were partially opposed and weakened in Refs.
\cite{smeared,discrete} where the assumption of ``starting with
$x$'' (i.e., of our working with the potentials which are local in
$x$) has been shown unfounded (because the value of the lower-case
$x$ is not observable) and misleading (because one need not give up
the unitarity in general). At the same time, the underlying, deep
and important conceptual problem of the possible existence of
suitable Hermitian -- non-Hermitian interfaces remained open.

An affirmative answer will be given in what follows. In order to
formulate the problem more clearly we will have to recall, in the
next section, a few well known aspects of forming a nontrivial
feasible contact and of a smooth transition between several versions
of quantum dynamics. In the subsequent sections we shall then point
out that a formal key to the realization of the project of
construction of the smooth interfaces lies in the properties of the
inner-product metric operators~$\Theta$ which have to degenerate
smoothly, in their turn, to the trivial limit $\Theta=I$.
Furthermore, in section~\ref{se6} several technical aspects of such
a general interface-construction recipe will be illustrated by an
elementary toy-model-Hamiltonian example admitting a non-numerical
and non-perturbative analytical treatment. Some of the possible
impacts upon quantum phenomenology will finally be mentioned in
section~\ref{se7}.


\section{Quantum dynamics in Schr\"{o}dinger picture}

During the birth of quantum theory its oldest (viz., the
Heisenberg's, ``matrix'') picture was quickly followed by the
Schr\"{o}dinger's ``wave-function'' formulation which proved less
intuitive but more economical \cite{Styer}. The conventional,
``textbook'' {\it alias\,} ``Hermitian'' Schr\"{o}dinger picture
(HSP, \cite{Messiah}) was later complemented by its
``non-Hermitian'' Schr\"{o}dinger picture (NHSP) alternative (cf.
the works by Freeman Dyson \cite{Dyson} or by nuclear physicists
\cite{Geyer}). In this direction the recent wave of new activities
was inspired by Carl Bender with coauthors \cite{Carl,BB}. The
emerging, more or less equivalent innovated versions of the NHSP
description of quantum dynamics were characterized as ``quantum
mechanics in pseudo-Hermitian representation'' \cite{ali} or as
``quantum mechanics in the Dyson's three-Hilbert-space formulation''
\cite{Dyson,SIGMA}, etc \cite{book}.

The availability of the two alternative representations of the laws
of quantum evolution in Schr\"{o}dinger picture inspired Hugh Jones
to ask the above-cited questions about the existence of an ``overlap
of their applicability'' in an ``interface'' \cite{London}. His
interest was predominantly paid to the scattering
\cite{preinterface} and his answers were discouraging
\cite{interface}. In papers~\cite{discrete,scatt} we opposed his
scepticism. We argued that the difficulties with the HSP - NHSP
interface may be attributed to the ultralocal, point-interaction
toy-model background of his methodical analysis. We introduced
certain weakly non-local interactions and via their constructive
description we reopened the possibility of practical realization of
a smooth transition between the Hermitian and non-Hermitian
theoretical treatment of scattering experiments.

Now we intend to return to the challenge of taking advantage of the
specific merits of {\em both} of the respective HSP and NHSP
representations inside their interface. We shall only pay attention
to the technically less complicated quantum systems with bound
states. Our old belief in the existence, phenomenological relevance
and, perhaps, even fundamental-theory usefulness of a domain of
coexistence of alternative Schr\"{o}dinger-picture descriptions of
quantum dynamics will be given an explicit formulation supported by
constructive arguments and complemented by elementary, analytically
solvable illustrative examples.

\subsection{The concept of hidden Hermiticity}


An optimal formulation of quantum theory is, obviously,
application-dependent \cite{Styer}. Still, the so called
Schr\"{o}dinger picture seems exceptional. Besides historical
reasons this is mainly due to the broad applicability as well as
maximal economy of the complete description of quantum evolution
using the single Schr\"{o}dinger equation
 \be
 {\rm i} \frac{d}{dt} \,\psi(t) = \mathfrak{h} \,\psi(t)\,,
 \ \ \ \ \ \ \ \ \ \ \
 \psi(t) \in {\cal H}^{(textbook)} \,.
 \label{HSP}
 \ee
Whenever the evolution is assumed unitary, the generator
$\mathfrak{h}$ (called Hamiltonian) must be, due to the Stone's
theorem \cite{Stone}, self-adjoint in ${\cal H}^{(textbook)}$,
 \be
 \mathfrak{h}=\mathfrak{h}^{(Hermitian)}=\mathfrak{h}^\dagger\,.
  \label{heha}
  \ee
Recently it has been emphasized that even in the unitary-evolution
scenario the latter Hamiltonian-Hermiticity constraint may be
omitted or, better, circumvented. The idea, dating back to Dyson
\cite{Dyson}, relies upon a suitable preconditioning of wave
functions. This induces the replacement of the ``Hermitian'',
lower-case Schr\"{o}dinger Eq.~(\ref{HSP}) + (\ref{heha}) by its
``non-Hermitian'' upper-case alternative
 \be
 {\rm i} \frac{d}{dt} \,\Psi(t) = H \,\Psi(t)\,,
 \ \ \ \ \ \ \
 \Psi(t) \in {\cal H}^{(unphysical)}\,,
 \ \ \ \ \ \ \
 \psi(t) = \Omega \, \Psi(t) \,.
  \label{NHSP}
 \ee
The preconditioning operator $\Omega$ is assumed invertible but
non-unitary, $\Omega^\dagger \Omega \neq I$ \cite{Geyer}. Thus, the
standard textbook version of the Schr\"{o}dinger picture splits into
its separate Hermitian and non-Hermitian versions (cf. influential
reviews \cite{ali,Carl} and/or mathematical commentaries in
\cite{book}).

The slightly amended forms of the Dyson's version of the NHSP
formalism proved successful in phenomenological applications, e.g.,
in nuclear physics \cite{Geyer}. As we already indicated, the
``non-Hermitian'' philosophy of Eq.~(\ref{NHSP}) was made widely
popular by Bender with coauthors \cite{Carl}. Its appeal seems to
result from the observation that the non-unitarity of $\Omega$ makes
the respective geometries in the two Hilbert spaces $ {\cal
H}^{(textbook)}$ and ${\cal H}^{(unphysical)}$ mutually
non-equivalent. As a consequence, the upper-case Hamiltonian $H$
acting in ${\cal H}^{(unphysical)}$ and entering the upgrade
(\ref{NHSP}) of Schr\"{o}dinger equation becomes manifestly
non-selfadjoint {\it alias\,} non-Hermitian in ${\cal
H}^{(unphysical)}$,
 \be
  H =H^{(non-Hermitian)}= \Omega^{-1} \mathfrak{h}\Omega
  \neq H^\dagger = \Theta H \Theta^{-1}
 \,,\ \ \ \ \ \ \Theta=\Omega^\dagger\Omega \neq I
  \,.
  \label{noheha}
   \ee
Still, it is obvious that {\em both\,} the NHSP version (\ref{NHSP})
of Schr\"{o}dinger equation {\em and\,} its HSP predecessor
(\ref{HSP}) represent {\em the same\,} quantum dynamics.

\subsection{The choice between the HSP and NHSP languages}

Several reviews in monograph \cite{book} may be recalled for an
extensive account of multiple highly nontrivial mathematical details
of the NHSP formalism. In applications, quantum physicists take it
for granted, nevertheless, that we have a choice between the {\em
two} alternative descriptions of the standard {\em unitary}
evolution of wave functions. People are already persuaded that the
basic mathematics of the HSP and NHSP constructions is correct and
that the two respective Schr\"{o}dinger equations are, for any
practical purposes, equally reliable.

The accepted abstract HSP - NHSP equivalence still does not mean
that the respective practical ranges of the two recipes are the
same. The preferences really {\em depend\,} very strongly on the
quantum system in question. Thus, the choice of the HSP language is
made whenever the corresponding self-adjoint Hamiltonian possesses
the most common form of superposition of a kinetic energy term with
a suitable {\em local-interaction\,} potential,
 \be
 \mathfrak{h}_{(local)}=-\frac{d^2}{dq^2} +
 \mathfrak{v}(q)=\mathfrak{h}^\dagger_{(local)}
 \,.
 \label{sei}
 \ee
Similarly, the recent impressive success of the NHSP
phenomenological models is almost exclusively related to the use of
the non-Hermitian local-interaction Hamiltonians
 \be
 H=-\frac{d^2}{dx^2} +
 W(x)\neq H^\dagger
 \label{selat}
 \ee
which are only required to possess the strictly real spectra of
energies \cite{ali,Carl}.

\subsection{The concept of the HSP - NHSP interface}

The two local-interaction operators (\ref{sei}) and (\ref{selat})
should be perceived as just the two illustrative elements of the two
respective general families ${\cal F}^{(H)}$ and ${\cal F}^{(NH)}$
of the eligible, i.e., practically tractable and sufficiently
user-friendly HSP and NHSP Hamiltonians. In a way influenced by this
exemplification one has a natural tendency to assume that the latter
two families are distinct and clearly separated, non-overlapping
\cite{Carl},
   \be
   {\cal F}^{(H)} \,\bigcap\,{\cal F}^{(NH)}
  =\emptyset\,.
   \label{ifexists}
   \ee
During the early stages of testing and weakening such an {\it a
priori} assumption, Hugh Jones \cite{London} introduced the concept
of an interface as a potentially non-empty set of Hamiltonians,
   \be
   {\cal F}^{(interface)}=
   {\cal F}^{(H)} \,\bigcap\,{\cal F}^{(NH)}
  \,.
   \label{deifexists}
   \ee
Basically, he had in mind a domain of a technically feasible and
phenomenologically consistent interchangeability of the two
pictures. He also outlined some of the basic features and possible
realizations of such a Hermitian/non-Hermitian interface in
Ref.~\cite{preinterface}. Incidentally, the continued study of the
problem made him more sceptical \cite{interface}. In a way based on
a detailed analysis of a schematic though, presumably, generic
toy-model local-interaction Hamiltonian
 \be
 H=H^{(non-Hermitian)}_{(local)}
 =-\frac{d^2}{dx^2} + V^{(Hermitian)}(x) +
  W^{(non-Hermitian)}(x)\,
  \label{model}
 \ee
he came to the conclusion that the merits of families ${\cal
F}^{(H)}$ and ${\cal F}^{(NH)}$ are really specific and that, in the
case of scattering at least, their respective domains of
applicability really lie far from each other, i.e., ${\cal
F}^{(interface)}_{(scattering)}
   =\emptyset$.
Even at the most favorable parameters and couplings, in his own
words, ``the physical picture [of scattering] changes drastically
when going from one picture to the other'' \cite{interface}.

In our first paper \cite{scatt} on the subject we pointed out that
the Jones' discouraging ``no-interface'' conclusions remain strongly
model-dependent. For another, weakly nonlocal choice of
$H^{(non-Hermitian)}_{(weakly\ nonlocal)}$ we encountered a much
less drastic effect of the interchange of the mathematically
equivalent Schr\"{o}dinger Eqs.~(\ref{HSP}) and (\ref{NHSP}) upon
the predicted physical outcome of the scattering (see also the
related footnote added in Ref.~\cite{interface}). In our subsequent
paper~\cite{discrete} we further amended the model and demonstrated
that in the context of scattering the overlaps ${\cal
F}^{(interface)}_{(scattering)}$ may be non-empty. We showed that
there may exist the sets of parameters for which the causality as
well as the unitarity would be guaranteed for {\em both\,} of the
Hamiltonians in Eqs.~(\ref{heha}) and (\ref{noheha}). Thus, the
Jones' ultimate recommendations of giving up the scattering models
in ${\cal F}^{(NH)}$ and/or of ``accepting \ldots the loss of
unitarity'' while treating any ``non-Hermitian scattering potential
as an effective one'' \cite{interface} may be re-qualified as
over-sceptical (cf. also \cite{ali}).

\section{Repulsion of eigenvalues\label{se2}}

The presentation of our results is to be preceded by a compact
summary of some of the key specific features of spectra in the
separate HSP and NHSP frameworks. This review may be found
complemented, in Appendix A, by a brief explanation why the NHSP
Hamiltonians $H$ which are {\em non-Hermitian} (though only in an
auxiliary, {\em unphysical\,} Hilbert space) still do generate the
{\em unitary\,} evolution (naturally, via wave functions in another,
non-equivalent, physical Hilbert space).


Quantum dynamics of the one-dimensional motion described by an
ordinary differential local-interaction Hamiltonian (\ref{sei}) is a
frequent target of conceptual analyses. These models stay safely
inside Hermitian class ${\cal F}^{(H)}$ but still, a brief summary
of some of their properties and simplifications will facilitate a
compact clarification of the purpose of our present study.

\subsection{Discrete coordinates}

The kinetic plus interaction structure of models (\ref{sei})
reflects their classical-physics origin. It may also facilitate the
study of bound states, say, by the perturbation-theory techniques
\cite{interface} and/or by the analytic-construction methods
\cite{Fluegge}. Still, for our present purposes it is rather
unfortunate that any transition to the hidden-Hermiticity language
of the alternative model-building family ${\cal F}^{(NH)}$ would be
counterproductive. One of the main obstacles of a hidden-Hermiticity
re-classification of model (\ref{sei}) is technical because the
associated Hamiltonians (\ref{noheha}) are, in general, strongly
non-local \cite{117}. Another, subtler mathematical obstacle may be
seen in the unbounded-operator nature of the kinetic energy $T =
-d^2/dx^2$ (see \cite{Geyer} for a thorough though still legible
explanation).

In Refs.~\cite{discrete,scatt} we proposed that one of the most
efficient resolutions of at least some of the latter problems might
be sought and found in the discretization of the coordinates. Thus,
one replaces the real line of $q \in (-\infty,\infty)$ by a discrete
lattice of grid points $q_j$ such that $q_j=q_0+h\,j$, with $j =
\ldots, -1,0,,1,\ldots$ and with any suitable constant $h>0$. This
leads to the kinetic energy represented by the difference-operator
Laplacean
 \be
  T
 = \left[ \begin {array}{rrrrr}
 \ddots&\ddots&&&
 \\\noalign{\medskip}\ddots
 &0&-1&&
 \\\noalign{\medskip}&-1&0&-1&
 \\\noalign{\medskip}&&-1&0&\ddots
 \\\noalign{\medskip}&&&\ddots&\ddots
 \end {array} \right]\,.
 \label{Tepr}
 \ee
In parallel, one can argue that the sparse-matrix structure of this
component of the Hamiltonian makes it very natural to replace also
the strictly local (i.e., diagonal-matrix) interaction
$\mathfrak{v}(q_j)$ by its weakly non-local tridiagonal-matrix
generalization~\cite{PRAbelow}.

\subsection{Elementary example}

Once we restrict our attention to the analysis of bound states, the
above-mentioned doubly infinite tridiagonal matrices
$\mathfrak{h}_{(weakly-local)}$ may be truncated yielding an $N$ by
$N$ matrix Hamiltonian. Let us assume here that the latter matrix
varies with a single real coupling strength $\epsilon$ and with a
single real parameter $\lambda$ modifying the interaction,
 \be
 \mathfrak{h}^{(\epsilon,\lambda)}= T +
 \epsilon\,\mathfrak{v}^{(\lambda)}
 =\left [\mathfrak{h}^{(\epsilon,\lambda)}
 \right ]^\dagger
 \,,\ \ \ \ \
 \epsilon, \lambda \in \mathbb{R}\,.
 \label{Hepr}
 \ee
This will enable us to assume that our parameters can vary,
typically, with time (i.e., $\epsilon=\epsilon(t)$ and/or
$\lambda=\lambda(t)$) and that, subsequently, also the energy levels
$E_n$ of our quantum system form a set which can, slowly or quickly,
vary. Thus, at a time of preparation $t=t_0$ of a Gedankenexperiment
the energy of our system may be selected as equal to one of the real
and time-dependent eigenvalues of our Hamiltonian $\mathfrak{h}$.
Naturally, the latter operator represents a quantum observable and
must be self-adjoint in the underlying physical Hilbert space ${\cal
H}^{(textbook)}$.

%

%
\begin{figure}[h]                     
\begin{center}                         
\epsfig{file=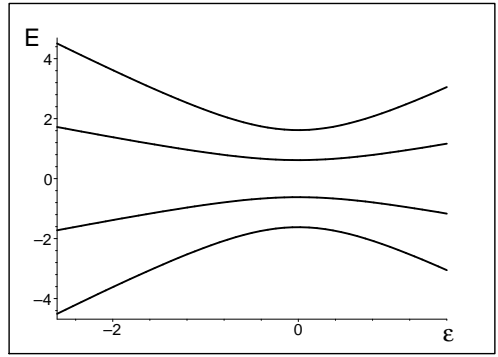,angle=270,width=0.4\textwidth}
\end{center}                         
\vspace{-2mm}\caption{The repulsion of the eigenvalues of the
Hermitian matrix (\ref{Hepr}) + (\ref{sTepr}) with $\lambda=1$ and
$N=4$ near $\epsilon=0$.
 \label{ee1wwww}}
\end{figure}


The first nontrivial tridiagonal matrix (\ref{Hepr}) with $N=4$ may
represent, e.g., a schematic quantum system with Hermitian-matrix
interaction
 \be
 \mathfrak{v}^{(\lambda)}
 = \left[ \begin {array}{rrrr} 0&{\rm i}&0&0
 \\\noalign{\medskip}-{\rm i}
 &0&{\rm i}\lambda&0
 \\\noalign{\medskip}0&-{\rm i}\lambda&0&{\rm i}
 \\\noalign{\medskip}0&0&-{\rm i}&0\end {array} \right]\,.
 \label{sTepr}
 \ee
The spectrum of energies may be then easily calculated and found
sampled in Fig.~\ref{ee1wwww}. The parameter-dependence of the
energies seems to be such that they avoid ``collisions''. As long as
we choose $\lambda=1$, i.e., Hamiltonian
 \be
 \mathfrak{h}^{(\epsilon,1)}
 =\left[ \begin {array}{cccc} 0&-1+i\epsilon&0&0\\\noalign{\medskip}-1-
 i\epsilon&0&-1+i\epsilon&0\\\noalign{\medskip}0&-1-i\epsilon&0&-1+i
 \epsilon\\\noalign{\medskip}0&0&-1-i\epsilon&0\end {array} \right]
 \label{mrepr}
 \ee
the quadruplet of the energy eigenvalues becomes available also in
the closed form
 \be
 E_{\pm,\pm}=\pm \frac{1}{2}\,\sqrt {\left (6 \pm 2\,\sqrt {5}\right )
 \left(1+{\epsilon}^{2}\right )}\,.
 \label{11}
 \ee
This formula explains not only the hyperbolic shapes of the curves
in Fig.~\ref{ee1wwww} but also their closest-approach values
$E_{\pm,+}\approx \pm 1.618033988$ and $E_{\pm,-}\approx \pm
0.6180339880$ at $\epsilon=0$.

The details of the generic avoided-crossing phenomenon are
model-dependent but an analogous observation will be made using {\em
any\,} Hermitian-matrix Hamiltonian. The explanation may be found in
the Kato's book \cite{Kato}. In essence, the Kato's mathematical
statement is that once a given matrix is self-adjoint {\it alias}
Hermitian, then in the generic case (i.e., without any additional
symmetries) an arbitrary pair of the eigenvalues can only merge at
the so called exceptional-point (EP) value of the parameter. In the
Hermitian diagonalizable (i.e., physical) cases these EP values are
all necessarily complex so that whenever the parameter remains real,
the distances between the separate real eigenvalues behave as if
controlled by a mutual ``repulsion'' \cite{footnote1}. From
Fig.~\ref{ee1wwww} we may then extract one of the key messages
mediated by the model, viz., the observation that the unitary
evolution is ``robust''. One may expect that whenever we need to
achieve an unavoided crossing of the eigenvalues, the more adequate
description of the phenomenon will be provided by the transition to
non-Hermitian Hamiltonians in ${\cal F}^{(NH)}$ for which the EP
values may be real.


\section{Attraction of eigenvalues\label{se3}}

The phenomenon of the existence of a minimal distance between the
energy levels of a Hermitian matrix is generic. After one tries to
move from family ${\cal F}^{(H)}$ to family ${\cal F}^{(NH)}$, the
robust nature of such an obstruction is lost. The reason lies in the
above-mentioned change of the geometry of the Hilbert spaces in
question. The resulting new freedom of models in ${\cal F}^{(NH)}$
may find applications, e.g., in an effective description of
non-unitarities in open quantum systems \cite{Samsonov} or, in
cosmology, in an elementary explanation of the possibility of a
consistently quantized Big Bang \cite{BiBa}.

\subsection{Local interactions}

One of the reasons of the recent turn of attention to the hiddenly
Hermitian local-interaction models (cf. their sample (\ref{model})
above) is that the mapping $H \to \mathfrak{h}$ of
Eq.~(\ref{noheha}) produces, in general, strongly nonlocal
generalizations of the conventional local Hamiltonians (\ref{sei}).
The same argument works in both directions and it enriched the scope
of the conventional quantum theory \cite{ali,Carl}. Several
impressive constructive illustrations of such a type of enrichment
of the class of the tractable quantum models (treating the direct
use of local-interaction models (\ref{selat}) as an important
extension of the applied quantum theory) may be found, e.g., in
Ref.~\cite{117}. One may conclude that the local-interaction nature
and constructive tractability of the alternative models
(\ref{model}) contained in class ${\cal F}^{(NH)}$ would render
their isospectral partners (\ref{heha}) non-local. Thus, some of the
weaker forms of the nonlocalities as sampled, e.g., in
Refs.~\cite{discrete,scatt} may be expected necessary for the
constructive search for the non-empty interfaces ${\cal
F}^{(H)}\bigcap{\cal F}^{(NH)}$.

\subsection{Weakly non-local interactions}

  \noindent
For the purposes of the most elementary though still sufficiently
rich illustration of some technical aspects of the transition from
${\cal F}^{(H)}$ to ${\cal F}^{(NH)}$ one may perform the
straightforward de-Hermitization of Eq.~(\ref{sTepr}). This yields
the two-parametric pencil of Hamiltonian matrices
 \be
 H^{(\eta,\lambda)}= T +
 \eta\,W^{(\lambda)}
 =
 \left[ \begin {array}{cccc}
 0&-1+\eta&0&0\\\noalign{\medskip}-1-\eta&0&-1+\eta\,\lambda&0
 \\\noalign{\medskip}0&-1-\eta\,\lambda&0&-1+\eta
 \\\noalign{\medskip}0&0&-1-\eta&0\end {array} \right]
 \neq
 \left [H^{(\eta,\lambda)}
 \right ]^\dagger\,
 \label{mrepro}
 \ee
characterized by a minimal, tridiagonal-matrix non-locality of their
interaction component. For real $\eta$ and $\lambda$ the related
energy spectra only remain real (i.e., observable and
phenomenologically meaningful) in certain physical parametric
intervals.


\begin{figure}[h]                     
\begin{center}                         
\epsfig{file=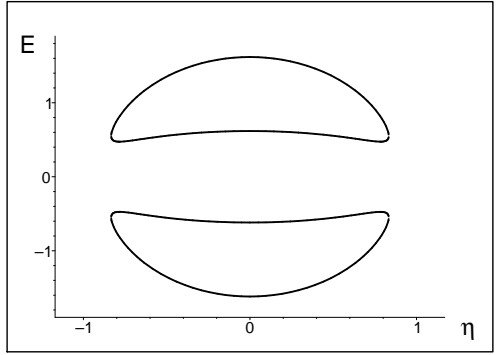,angle=270,width=0.4\textwidth}
\end{center}                         
\vspace{-2mm}\caption{The attraction (followed by the loss of
reality) in the case of the eigenvalues of the manifestly
non-Hermitian matrix (\ref{mrepro}) at $\lambda=6/5$.
 \label{ee2wb}}
\end{figure}

The simplicity of our toy model (\ref{mrepro}) enables us to
illustrate the latter statement by recalling the explicit formula
for the eigenvalues,
 \be
 E_{(\pm,\pm)}(\epsilon,\lambda)=\pm \frac{1}{\sqrt{2}}\,\sqrt {3-\left ({
\lambda}^{2}+2\right )\,{\epsilon}^{2}\pm \sqrt {
  \left[ 5-\left ({\lambda}^{2}+4\right )\,{\epsilon}^{2}
 \right]\left(
1-\lambda^2\,\epsilon^2 \right) }}\,\ .
 \ee
%
%
The knowledge of this formula enables us to separate the interval of
the interaction-controlling parameters $\lambda$ into three
qualitatively different subintervals.

\subsubsection{$\lambda > 1$ (strong non-Hermiticities)}

 \noindent
The first,  $\lambda>1$ sample of the energy spectrum is displayed
here in Fig.~\ref{ee2wb}. The picture shows that at the two real
exceptional points $\eta=\eta^{(EP)}$ such that
$|\eta|=|\eta^{(EP)}|< 1$ the (real-energy) quadruplets of energies
degenerate and, subsequently, acquire imaginary components. These
complexifications proceed pairwise, i.e., our four-level model
effectively decays into two almost independent, weakly coupled
two-level systems. The full descriptive wealth of our model will
only manifest itself at the smaller values of $\lambda$.


%
\begin{figure}[h]                     
\begin{center}                         
\epsfig{file=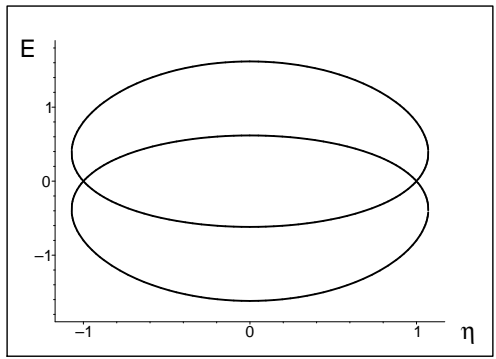,angle=270,width=0.4\textwidth}
\end{center}                         
\vspace{-2mm}\caption{Same as Fig.~\ref{ee2wb}, with smaller
$\lambda=3/5$.
 \label{ee2wbc}}
\end{figure}

\subsubsection{$\lambda < 1$ (weak non-Hermiticities)}

 \noindent
In Fig.~\ref{ee2wbc} using a smaller $\lambda<1$ a much more
interesting scenario is displayed in which all of the four energy
levels are mutually attracted. Firstly we notice that the
complexifications of the eigenvalues occur at the EP values
$\eta^{(EP)}_{(first\ kind)}$ which are ``large'', i.e., such that
$|\eta^{(EP)}_{(first\ kind)}|>1$. The domain of the observability
of the energies is larger than interval $(-1,1)$. Still, the latter
interval has natural boundaries because of the emergence of the
other two EP degeneracies at $\eta^{(EP)}_{(second\ kind)}=\pm 1$.
These new singularities are characterized by the unavoided level
crossings without a complexification. Their occurrence splits the
interval of $\eta$ into separate subintervals. The consequences for
the quantum phenomenology are remarkable, e.g., for the reasons
which were discussed, recently, in \cite{Denis}

%
\begin{figure}[h]                     
\begin{center}                         
\epsfig{file=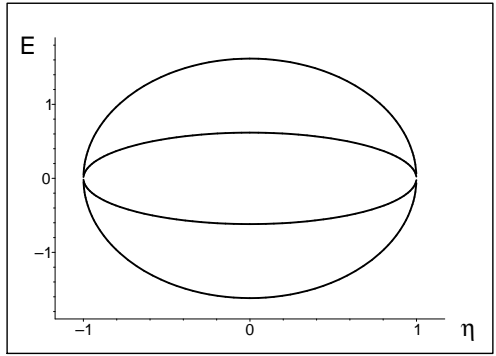,angle=270,width=0.4\textwidth}
\end{center}                         
\vspace{-2mm}\caption{The confluence of exceptional points at
$\lambda = 1$.
 \label{ee2w}}
\end{figure}

\subsubsection{$\lambda = 1$ (the instant of degeneracy)}

The shared boundary between the two dynamical regimes is characterized by
Fig.~\ref{ee2w}. The algebraic representation of the $\lambda = 1$
spectrum is elementary,
%
%
 \be
 E_{\pm,\pm}=\pm \frac{1}{2}\,\sqrt {\left (6 \pm 2\,\sqrt {5}\right )
 \left(1-{\eta}^{2}\right )}\,.
 \label{14}
 \ee
%
The formula may be read as an analytic continuation of
Eq.~(\ref{11}).

%

\section{The model with interface \label{se6}}

\subsection{Hilbert-space metric}

In comparison with the conventional textbook family ${\cal
F}^{(H)}$, the practical use of the non-Hermitian phenomenological
Hamiltonians in ${\cal F}^{(NH)}$ is certainly much more difficult.
One of the key complications is to be seen in the (in general,
non-unique) reconstruction of the metric from the given observables
or, in the simplest case, from Hamiltonian $H$.

The ambiguity of the reconstruction may be illustrated by the
insertion of our two-parametric toy-model $N=4$ Hamiltonian
$X=H^{(\eta,\lambda)}$ of Eq.~(\ref{mrepro}) in the
quasi-Hermiticity constraint (\ref{die}) in Appendix A interpreted
as an implicit definition of
$\Theta^{(\eta,\lambda)}=\Theta(H^{(\eta,\lambda)})$. After a
tedious but straightforward algebra one obtains the general result
 \be
 \Theta^{(\eta,\lambda)}_{(c,d,f,g)}=
 \left[ \begin {array}{cccc} {\it A}(f,c)&{\frac { \left( g-d \right)
  \left( 1+\eta \right) }{1-\eta\,\lambda}}&{\frac {{\it {c}}}{1-\eta}}&
 d
 \\\noalign{\medskip}{\frac { \left( g-d \right)  \left( 1+\eta
 \right) }{1-\eta\,\lambda}}&{\frac {{\it {f}}}{1-\eta\,\lambda}}&g&{
 \frac {{\it {c}}}{1+\eta}}
 \\\noalign{\medskip}{\frac {{\it
 {c}}}{1-\eta} }&g&{\frac {{\it {f}}}{1+\eta\,\lambda}}&{\frac {
 \left( g-d \right)
 \left( 1-\eta \right) }{1+\eta\,\lambda}}
 \\\noalign{\medskip}d&{
 \frac {{\it {c}}}{1+\eta}}&{\frac { \left( g-d \right)  \left( 1-\eta
 \right) }{1+\eta\,\lambda}}&{\it F}(f,c)
 \end {array} \right]
 \label{where}
 \ee
where
 \be
 {\it A}(f,c)= {\frac {{\it {f}}-{\it {f}}\,{\eta}^{2}-{\it {c}}+{\it
{c}}\,{\eta}^{2}{ \lambda}^{2}}{\left( 1-\eta \right) ^{2} \left(
1-\eta\,\lambda \right)}}
 \ee
and
 \be
 {\it C}(f,c)={\frac {{\it {f}}-{\it {f}}\,{\eta}^{2}-{\it {c}}+{\it {c}}\,{\eta}^{2}{
\lambda}^{2}}{\left( 1+\eta \right) ^{2} \left( 1+\eta\,\lambda
\right)}}\,.
 \ee
Thus, one can summarize that unless we add more requirements, the
specification of the mere Hamiltonian leads to the four-parametric
family of the inner-product metric-operators (\ref{where}).
Obviously, this opens the possibility of the choice of the
additional observables which would have to satisfy Eq.~(\ref{die})
and, thereby, restrict the freedom in our choice of the parameters
$c,d,f$ and $g$.

%
%


One of the possible formal definitions of an ``interface'' between
the alternative descriptions (\ref{HSP}) and (\ref{NHSP}) of a
quantum system may be based on the presence of a variable parameter
or parameters (say, of a real $\sigma \in (-\infty,\infty)$) such
that $\mathfrak{h}=\mathfrak{h}(\sigma)$ and $H=H(\sigma)$. One may
then reveal that there exists a point $\sigma_0$ or a non-empty
closed vicinity $I_0=(\sigma_-,\sigma_+)$ of this point such that
the formally equivalent Schr\"{o}dinger Eqs.~(\ref{HSP}) and
(\ref{NHSP}) are also more or less equally user-friendly when
$\sigma \in I_0$. Naturally, such a concept will make sense when
just the solution of one of the Schr\"{o}dinger equations remains
feasible and practically useful far from $\sigma_0$.

Whenever one tries to treat $\sigma$ as a function of time, a number
of technical complications immediately emerges (the most recent
account of some of them may be found in \cite{NIP}). One has to
assume, therefore, that the time-variation of $\sigma$ as well as
the $\sigma-$variation of the Hamiltonians remains sufficiently
slow, i.e., so slow that the corresponding time-derivatives of
$\sigma$ and the $\sigma-$derivatives of the Hamiltonians remain
negligible. Under these assumptions, the passage of certain quantum
systems through their respective HSP - NHSP interfaces can be shown
possible.

\subsection{Illustrative Hamiltonian}

One of the most straightforward implementations of the above idea
may be based on the identification of the above-introduced parameter
$\sigma$ with the parameter $\epsilon$ of Eq.~(\ref{Hepr}) (and,
say, of Fig.~\ref{ee1wwww}) along the negative real half-axis, and
with the parameter $\eta$ of Eq.~(\ref{mrepro}) (and of
Fig.~\ref{ee2w}) along the positive real half-axis. In such an
arrangement the interval of a large and negative $\sigma \ll -1$
will be the domain in which the use of the non-Hermitian picture
${\cal F}^{(NH)}$ (with any nontrivial metric) would prove
absolutely useless. In parallel, any attempt of working with the
Hermitian picture ${\cal F}^{(H)}$ will necessarily fail close to
$\sigma \approx +1$ and further to the right. At the same time, in
practically any interval of the positive $\sigma = \eta \in
I_0=(0,\sigma_+)$ with $\sigma_+ < 1$ we would be able to work, more
or less equally easily, with both of the non-Hermitian and Hermitian
versions of the matrix.

The main advantage of the work in simultaneous pictures, i.e., with
the Hamiltonian matrix defined in ${\cal F}^{(interface)}$ may be
seen in the smoothness of the transitions to both of the neighboring
pictures ${\cal F}^{(H)}$ and ${\cal F}^{(NH)}$. This smoothness is
nontrivial because the respective behaviors of the quantum system in
question will be different, in spite of the unified definition of
the dynamics.
Thus, once we set
 \be
 H^{(unified)}= \left[ \begin {array}{cccc} 0&-1+\gamma(\tau)&0&0
  \\\noalign{\medskip}-1-\gamma(\tau)&0
 &-1+\gamma(\tau)&0
 \\\noalign{\medskip}0&-1-\gamma(\tau)&0&-1+\gamma(\tau)
 \\\noalign{\medskip}0
 &0&-1-\gamma(\tau)&0\end {array} \right]
 \label{mreprod}
 \ee
with
 \be
 \gamma(\tau)
 =\sqrt{\tau^2 \cdot {\rm sign}\ \tau}
 =\sqrt{\tau \cdot |\tau|}=
 \left \{
 \begin{array}{ll}
 {\rm i}\tau\,, \ \ &\tau < 0\,,\\
 \tau\,,&\tau \geq 0
 \ea
 \right .
 \ee
we will be able to interpolate, smoothly, between the eigenvalue
repulsion to the left and the eigenvalue attraction to the right
(see Fig.~\ref{ee3w}).

%
\begin{figure}[h]                     
\begin{center}                         
\epsfig{file=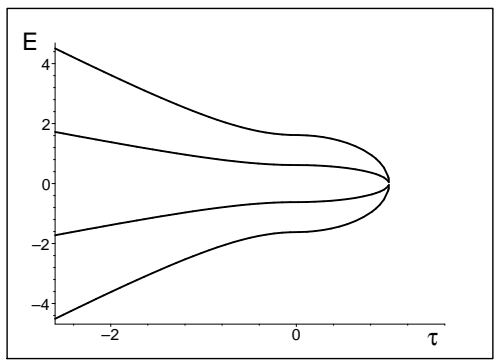,angle=270,width=0.4\textwidth}
\end{center}                         
\vspace{-2mm}\caption{Monotonic convergence of eigenvalues of matrix
(\ref{mreprod}) with the growth of $\tau$.
 \label{ee3w}}
\end{figure}

In addition, one may also appreciate the asymmetry of the spectrum.
In the purely phenomenological setting it could be interpreted,
e.g., as a transition from the conventional and robust dynamical
regime to the emergence of an instability and collapse at positive
$\tau = 1$. Marginally, let us also note that our choice of notation
is indicative because $\tau$ might have been perceived as a time
variable, in an adiabatic regime at least \cite{NIP}.
%
%
Another marginal comment is that at  $\tau > \tau^{(EP)}=1$, i.e.,
at $\tau =\sqrt{1+\varrho^2}> 1$, the eigenvalues form the two
purely imaginary complex-conjugate pairs
 \be
 \pm \frac{{\rm i}\varrho}{2}\,\sqrt {\left (6 \pm 2\,\sqrt {5}\right )
 }\approx
 \left \{
 \begin{array}{ll}
 \pm 1.618033988\ {\rm i}\varrho\\
 \pm 0.6180339880\ {\rm i}\varrho
 \ea
 \right .\,.
 \label{twopa}
 \ee
%
%
%
%
%
%
%
%
%

%
%

%
\begin{figure}[h]                     
\begin{center}                         
\epsfig{file=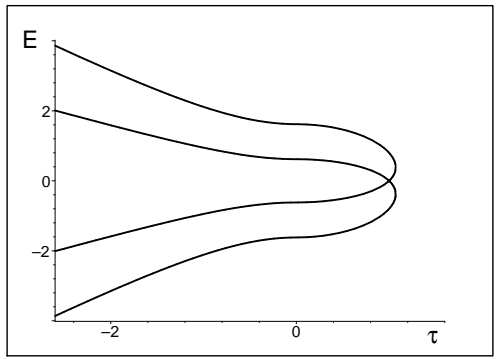,angle=270,width=0.4\textwidth}
\end{center}                         
\vspace{-2mm}\caption{Same as Fig.~\ref{ee3w} but with a slightly
smaller $\lambda=3/5$.
 \label{ee5w}}
\end{figure}

In the light of our preceding analysis it is not too surprising that
for the larger values of $\lambda>1$ the simultaneous
complexification of the eigenvalues would occur at a slightly
smaller EP singularity $\tau^{(EP)}<1$ and that the model would
effectively decay into the two two-level subsystems.
Such an observation might be contrasted with the more interesting
spectral pattern obtained at $\lambda=3/5$ and displayed in
Fig.~\ref{ee5w}.

\subsection{The interface-compatible metrics}

The physical interpretation of the
parameter $\sigma$ need not be specified at all. Its interface
values $\sigma_0 \in I_0$  might mark a critical time or
the position of a spatial
boundary or a critical value of the
strength of influence of an environment,
etc.

In our present illustrative model the
specification of the left boundary point $\sigma_-=0$ is unique
because of the natural choice of $\Theta=I$ along the whole negative
half-axis of $\sigma$. In contrast, our choice of the right boundary
point $\sigma_+<1$ remains variable because we always have $\Theta
\neq I$ for all of the positive physical values of $\sigma$.


We have to match the Hermitian choice of $\Theta=I$ valid at the
negative half-axis of $\sigma\leq 0$ to the hidden-Hermiticity
choice of $\Theta \neq I$ at the small and positive $\sigma> 0$. We
may recall formula (\ref{where}) and deduce that
 \be
 \lim_{\eta \to 0}\,\Theta^{(\eta,\lambda)}_{(c,d,f,g)}=
 \left[ \begin {array}{cccc} {\it {f}}-{\it c}&g-d&{\it c}&d
 \\\noalign{\medskip}g-d&{\it {f}}&g&{\it c}\\\noalign{\medskip}{\it c
 }&g&{\it {f}}&g-d\\\noalign{\medskip}d&{\it c}&g-d&{\it {f}}-{\it
 c}
 \end {array} \right]\,.
 \ee
Even if we admit that the values of the parameters in the metric may
be $\eta-$dependent, $c=c(\eta),d=d(\eta),f=f(\eta)$ and
$g=g(\eta)$, we must demand that  $d(0)=c(0)=g(0) = 0$ and
normalize, say, $f(0)=1$. This yields the metric which is diagonal
at $\eta=0$ and which remains diagonal after we require that the
parameters remain constant, $\eta-$independent. The elements forming
the diagonal of such a special Hilbert-space metric $\Theta$ read
 \be
 \left \{
 \frac{1+\eta}{(1-\eta)(1-\eta\lambda)},\frac{1}{1-\eta\lambda},
 \frac{1}{1+\eta\lambda},\frac{1-\eta}{(1+\eta)(1+\eta\lambda)}
 \right \}\ .
 \label{lala}
 \ee
In Fig.~\ref{ee7wb} we may see the coincidence of these elements in
the limit $\eta \to 0$, demonstrating the smooth variation of the
metric $\Theta$ in the both-sided vicinity of $\eta=0$.

%
%
\begin{figure}[h]                     
\begin{center}                         
\epsfig{file=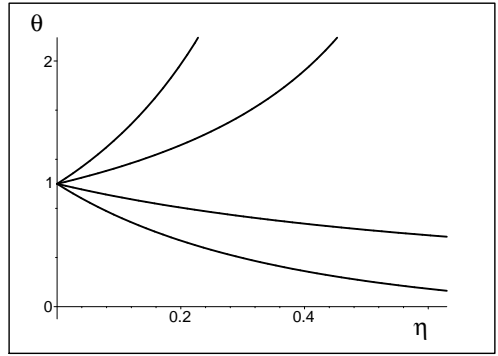,angle=270,width=0.4\textwidth}
\end{center}                         
\vspace{-2mm}\caption{The $\eta-$dependence of the eigenvalues of
the metric $\Theta$ of Eq.~(\ref{lala}) at $\lambda=6/5$.
 \label{ee7wb}}
\end{figure}

The construction of the kinetic-energy part $T$ of all of our toy
model matrix Hamiltonians $H$ with $N=4$ was based on the assumption
that there exist coordinates $q$ forming a spatial grid-point
lattice. In the present context this means that once the metric
(\ref{lala}) remains diagonal, in an interval of small
$\sigma=\eta>0$ at least, we may conclude that the
strong-non-locality effects as caused by the metric $\Theta$ and
observed, say, in Refs.~\cite{interface,117} are absent here. In
this sense, our present model shares the weak-nonlocality merits of
its predecessors in Refs.~\cite{discrete,scatt}.
%
%

\begin{figure}[h]                     
\begin{center}                         
\epsfig{file=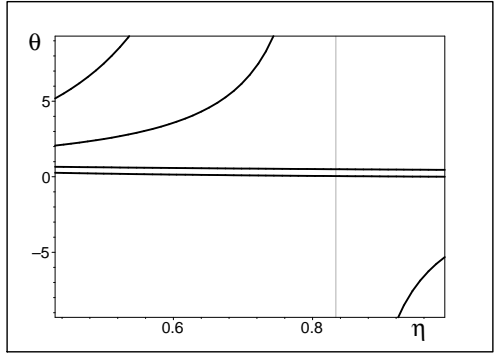,angle=270,width=0.4\textwidth}
\end{center}                         
\vspace{-2mm}\caption{The prolongation of Fig.~\ref{ee7wb} beyond
$\eta^{(EP)}=5/6$.
 \label{ee7wbw}}
\end{figure}

Our diagonal metric remains positive and invertible, at the
sufficiently small $\eta$ at least. Naturally, it also has the
EP-related singularities at $\eta^{(EP)}_{(first\ kind)}=\pm
1/\lambda$. Their occurrence and $\lambda-$dependence is illustrated
here in Fig.~\ref{ee7wbw}. Naturally, for $\lambda<1$ there emerge
also the singularities at $\eta^{(EP)}_{(second\ kind)}=\pm 1$ (see
the dedicated Ref.~\cite{Denis} for a more thorough explanation of
this terminology).

\section{Summary and conclusions \label{se7}}

In the conventional applications to quantum theory, the description
of the unitary evolution of a given system ${\cal S}$ need not
necessarily be performed in Schr\"{o}dinger picture (cf., e.g., the
compact review of its eight eligible alternatives in \cite{Styer}).
Naturally, once people decide to prefer the work in Schr\"{o}dinger
picture, they usually recall the Stone's theorem \cite{Stone} and
conclude that the Hamiltonian (i.e., in our present notation,
operator $\mathfrak{h} \in {\cal F}^{(H)}$ acting in the
conventional Hilbert space ${\cal H}^{(textbook)}$) must necessarily
be Hermitian (for the sake of brevity we spoke here about the HSP
realization of Schr\"{o}dinger picture).

Along a complementary, different line of thinking which dates back
to Dyson \cite{Dyson} and which recently climaxed with Bender
\cite{Carl} and Mostafazadeh \cite{ali}, the community of physicists
already accepted the consistency of the alternative, NHSP
realization of the same Schr\"{o}dinger picture. In the NHSP version
and language the Hamiltonian (i.e., the upper-case operator $H \in
{\cal F}^{(NH)}$ with real spectrum) is naturally self-adjoint in
the physical Hilbert space ${\cal H}^{(redefined)}$ which is,
unfortunately, highly unconventional. The same operator $H$ only
{\em appears\,} manifestly non-Hermitian in the other, auxiliary,
``redundant'' Hilbert space ${\cal H}^{(unphysical)}$ which is, by
assumption, ``the friendliest'' one.

It is unfortunate that the latter, historically developed
terminology is so confusing. This is one of the explanations why the
methodically important question of the possible HSP/NHSP overlap of
applicability has not yet been properly addressed and clarified in
the literature. In our present paper we filled the gap by showing
that such an overlap (called, by Jones \cite{interface}, an
``interface'') may exist. We also emphasized that the construction
of the interface should start from the upper-case (and, typically,
one-parametric) family of the hiddenly Hermitian NHSP Hamiltonians
operators $H =H(\sigma) \in {\cal F}^{(NH)}$ and that it has to be
based on the analysis of the related family of the Hermitizing
metric operators $\Theta=\Theta[H(\sigma)]$.

In such a framework one can conclude that Hamiltonian $H =H(\sigma)$
with $\sigma \in (\sigma_-,\sigma_+)$ can be perceived as an element
of an HSP/NHSP overlap ${\cal F}^{(interface)}\neq \emptyset$,
provided only that the Hermitizing metric operator at our disposal
(i.e., operator $\Theta=\Theta[H(\sigma)]$) is such that
 \be
 \lim_{\sigma\to\sigma_-}\Theta[H(\sigma)]=I\,.
 \ee
In other words, once we have $H(\sigma_-)=H^\dagger(\sigma_-)$, we
may now introduce the quantum Hamiltonians $\mathfrak{h}(\sigma)$
(which lie, by construction, in ${\cal F}^{(H)}$) in such a way that
they are connected with $H(\sigma)$ (i.e., defined) by relation
(\ref{noheha}) at $\sigma\in (\sigma_-,\sigma_+)$ while their
definition may be continued to $\sigma<\sigma_-$ arbitrarily (e.g.,
by the most straightforward constant-operator prescription
$\mathfrak{h}(\sigma)=H(\sigma_-)$).

The lower boundary $\sigma_-0$ of the interval of the
interface-compatible parameters carries an immediate physical
meaning of a point of transition from the HSP eigenvalue repulsion
regime (guaranteeing the robust reality of the spectrum) to the NHSP
eigenvalue attraction (and, possibly, complexification). Via na
elementary illustrative example we demonstrated that the resulting
``mixed'' dynamics could enrich the current phenomenological
considerations in quantum theory. Naturally, this is a task for
future research because our present, methodically motivated and
analytically solvable example is only too schematic for such a
purpose.

\section*{Acknowledgments}
The work was supported by the GA\v{C}R Grant Nr. 16-22945S.

\newpage

\section*{Appendix A. Unitary evolution via non-Hermitian $H$}



\subsection*{A. 1. The third Hilbert space}

Strictly speaking, the real spectra of eigenvalues of $H \neq
H^\dagger$ as well as of any other operator $X \neq X^\dagger$ of
the observable characterizing the quantum system in question cannot
be assigned any immediate physical meaning because the underlying
Hilbert space ${\cal H}^{(unphysical)}$ is, by definition, just
auxiliary and ``incorrect''. The ``correct'' meaning of the
observables can only be established in the ``correct'' Hilbert space
${\cal H}^{(textbook)}$. Whenever needed, any experimental
prediction may be reconstructed using the correspondences
 $
 \psi(t) = \Omega\,\Psi(t)$,
 $\mathfrak{h}=\Omega\,H\,\Omega^{-1}\,
 $ and
 $\mathfrak{x}=\Omega\,X\,\Omega^{-1}\,
 $.

One of the benefits of the NHSP representation is that in the
generic stationary case the full knowledge of the Dyson's operator
$\Omega$ is not necessary. What controls the predictions are just
the mean values of the operators of observables. For them, the
translations of the relevant formulae from ${\cal H}^{(textbook)}$
to ${\cal H}^{(unphysical)}$ may be shown to contain only the so
called Hilbert-space metric, i.e., the Dyson-map product
$\Theta=\Omega^\dagger \Omega$ (see, e.g., \cite{Geyer} for more
details). This implies that in a close parallel to
Eq.~(\ref{noheha}), all of the observables of a system in question
may be represented by the diagonalizable operators $X \neq
X^\dagger$ with real spectra which only have to satisfy the
generalized Hermiticity relation
 \be
 X^\dagger \Theta=\Theta\,X\,.
 \label{die}
 \ee
Any Hilbert space metric $\Theta$ which is ``mathematically
acceptable'' (see \cite{ali} for details) may be interpreted as
redefining the inner product in ${\cal H}^{(unphysical)}$. This
redefinition of the inner product may be re-read as a redefinition
of the Hilbert space itself,
 \be
 {\cal H}^{(unphysical)}\to {\cal
 H}^{(redefined)}\,.
  \ee
By construction, the new space becomes unitarily equivalent to
${\cal H}^{(textbook)}$. This means that we may re-interpret
Hamiltonians $H$ (sampled by Eq.~(\ref{model}) and non-Hermitian in
auxiliary ${\cal H}^{(unphysical)}$) as self-adjoint in the new
Hilbert space ${\cal H}^{(redefined)}$. Thus, using the notation of
Ref.~\cite{SIGMA} we may write $H=H^\ddagger$, with the definition
of $H^\ddagger= \Theta^{-1}H^\dagger\Theta$ being deduced from
Eq.~(\ref{noheha}) above.

In opposite direction our quantum-model-building may start from a
given $N-$plet $\{X_n\}$ of candidates for the observables. As long
as all of these operators (defined in ${\cal H}^{(unphysical)}$)
must satisfy the respective hidden Hermiticity condition
(\ref{die}), there must exist a metric candidate
$\Theta=\Theta(X_1,\ldots, X_N)$ compatible with all of these
hidden-Hermiticity conditions. Thus, the metric need not exist at
all (see an example in \cite{arabky}). If it does exist, it may be
either ambiguous (see an example in \cite{117}) or unique (see,
e.g., a large number of examples in \cite{Carl}).

\subsection*{A. 2. Physical inner products}

The non-Hermiticity property of operators might cause complications
in calculations. Also the assumptions of the user-friendliness of
${\cal H}^{(unphysical)}$ and/or of $H(t)$ seem highly nontrivial.
On the level of theory one must keep in mind that the new,
friendlier Hilbert space is, by itself, merely auxiliary and
unphysical. In principle, a return to ${\cal H}^{(textbook)}$ is
needed whenever experiment-related predictions are asked for. Still,
whenever the structure of such a space and/or of the observables
(defined in this space and sampled by Hamiltonian $\mathfrak{h}$)
appear prohibitively complicated, the evaluation of the predictions
of the theory is to be made also directly in ${\cal
H}^{(unphysical)}$. Due care must only be paid to the insertions of
the metric operator $\Theta=\Omega^\dagger\Omega \neq I$ (i.e., to
the amendments of the inner products) whenever applicable
\cite{ali}.

%
\begin{figure}[h]                     
\begin{center}                         
\epsfig{file=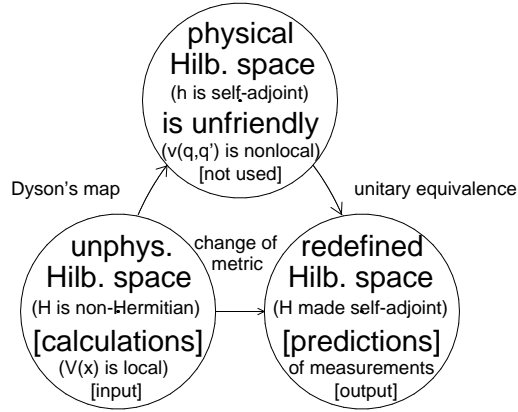,angle=270,width=0.6\textwidth}
\end{center}                         
\vspace{-2mm}\caption{The three-Hilbert-space representation
pattern.
 \label{e0wwww}}
\end{figure}

People do not always notice that after the latter amendment of the
inner product our auxiliary Hilbert space ${\cal H}^{(unphysical)}$
becomes redefined and converted into another, third Hilbert space
${\cal H}^{(redefined)}$ which is, by construction, physical, i.e.,
unitarily equivalent to ${\cal H}^{(textbook)}$. Thus, whenever we
start from Eq.~(\ref{NHSP}), the quantum system in question becomes
{\em simultaneously\,} represented in a {\em triplet\,} of Hilbert
spaces (the pattern is displayed in Fig.~\ref{e0wwww}).

Naturally, the Stone's theorem does not get violated due to the
one-to-one, $\Omega-$mediated correspondence between $H$ and
$\mathfrak{h}$. Due to the property $\Omega^\dagger\Omega=
\Theta\neq I$ of the Dyson's non-unitary mappings, the Hermiticity
of the conventional Hamiltonian $\mathfrak{h}$ in the physical space
${\cal H}^{(textbook)}$ becomes replaced, in the auxiliary and
manifestly unphysical Hilbert space ${\cal H}^{(unphysical)}$, by
the hidden Hermiticity {\it alias} $\Theta-$pseudo-Hermiticity
\cite{ali} property
 $
 H = \Theta^{-1} H^\dagger\Theta\
 $
of the upper-case non-Hermitian Hamiltonian with real spectrum (cf.
Eq.~(\ref{NHSP}) above). In the related literature one can also read
about the closely related concepts of quasi-Hermiticity (see
\cite{Geyer}), unbroken ${\cal PT}-$symmetry \cite{Carl} or
crypto-Hermiticity \cite{SIGMA,Smilga} of $H$ and/or, last but not
least, about the quasi-similarity between $H$ and $H^\dagger$
\cite{ATbook}.


\subsection*{A. 3. The Hermitian-theory point of view}

Technically, it is usually easier to work with the elements of the
``Hermitian'' family ${\cal F}^{(H)}$ comprising the traditional
quantum systems and the traditional textbook self-adjoint
Hamiltonians $\mathfrak{h}=\mathfrak{h}^\dagger$. Dyson \cite{Dyson}
merely proposed that sometimes, it may still make sense to make use
of the other, innovative family ${\cal F}^{(NH)}$ which works with
the ``non-Hermitian'' Schr\"{o}dinger Eqs.~(\ref{NHSP}). Certainly,
the latter family is not small.  {\it Pars pro toto\,} it contains
Hamiltonians of relativistic quantum mechanics \cite{alikg,Smejkal},
the well known ${\cal PT}-$symmetric imaginary cubic oscillator
\cite{Caliceti,DB} (which appears, after a more detailed scrutiny,
strongly non-local \cite{117,SK}), its power-law generalizations
\cite{BB,1nadeltob} as well as exactly solvable models \cite{ES},
models with methodical relevance in the context of supersymmetry
\cite{susy}, realistic and computation-friendly interacting-boson
models of heavy nuclei \cite{Geyer}, benchmark candidates for
classification of quantum catastrophes \cite{maximal}, etc.

In the majority of the above-listed models defined in ${\cal
F}^{(NH)}$ one may still keep in mind that their physical contents
can always be sought in their equivalence to the partner
Hamiltonians (and/or other observables) in ${\cal F}^{(H)}$. Thus,
the use of the less usual representation in ${\cal F}^{(NH)}$ is
treated as a mere technical trick.

The main argument against the latter, fairly widespread point of
view may be formulated as an objection against the over-intimate,
history-produced relationship between the way of our thinking in
classical physics and the related production of the ``conventional''
quantum models in ${\cal F}^{(H)}$ by the techniques of the so
called ``quantization''. In principle, we should have been much more
humble, taking rather the classical world as a result of making its
quantum picture ``de-quantized''~\cite{Lane}.


\begin{thebibliography}{99}
\bibitem{footnote1}
A deeper analysis may be found in the Kato's book \cite{Kato}.
\end{thebibliography}
\end{document}